%% file: main.tex
\documentclass[onecolumn]{IEEEtran}
\usepackage[T1]{fontenc}
\usepackage[inline]{enumitem}
\usepackage[latin9]{inputenc}
\usepackage[normalem]{ulem}
\usepackage{amsthm}
\usepackage{amsmath}
\usepackage{amssymb}
\usepackage{algpseudocode}
\usepackage{bm}
\usepackage{breakcites}
\usepackage{apacite}
\usepackage{color}
\usepackage{colortbl}
\usepackage{float}
\usepackage{graphicx}
\usepackage{hhline}
\usepackage{listings}
\usepackage{mathrsfs}
\usepackage{multicol}
\usepackage{multirow}
\usepackage{url}
\usepackage{pgfplots}
\usepackage{stfloats}
\usepackage{subcaption}

\makeatletter

\floatstyle{ruled}
\newfloat{algorithm}{tbp}{loa}
\providecommand{\algorithmname}{Algorithm}
\floatname{algorithm}{\protect\algorithmname}
\usepackage{tikz}
\usetikzlibrary{calc}

\theoremstyle{plain}
\newtheorem{lem}{\protect\lemmaname}
\theoremstyle{plain}
\newtheorem{cor}{\protect\corollaryname}
\theoremstyle{plain}
\newtheorem{prop}{\protect\propositionname}
\theoremstyle{plain}
\newtheorem{thm}{\protect\theoremname}

\usepackage[acronym]{glossaries}
\newcommand{\newac}{\newacronym}
\newcommand{\ac}{\gls}
\newcommand{\Ac}{\Gls}
\newcommand{\acpl}{\glspl}
\newcommand{\Acpl}{\Glspl}
\input{reference/acronym}

\makeatother

\providecommand{\corollaryname}{Corollary}
\providecommand{\lemmaname}{Lemma}
\providecommand{\propositionname}{Proposition}
\providecommand{\theoremname}{Theorem}

\newcommand{\revadd}[1]{#1}
\definecolor{mycolor1}{rgb}{0.494117647058824,0.184313725490196,0.556862745098039}
\definecolor{mycolor2}{rgb}{0.466666666666667,0.674509803921569,0.188235294117647}
\definecolor{mycolor3}{rgb}{0.301960784313725,0.745098039215686,0.933333333333333}
\definecolor{mycolor4}{rgb}{0.929411764705882,0.694117647058824,0.125490196078431}
\definecolor{mycolor5}{rgb}{0.635294117647059,0.078431372549020,0.184313725490196}
\definecolor{mycolor6}{rgb}{0.8500,0.3250,0.0980}

\title{Extending RAIM with a Gaussian Mixture of Opportunistic Information}

\author{
    Wenjie~Liu and Panos~Papadimitratos%
    \vspace{1mm} \\%
    \textit{Networked Systems Security Group, KTH Royal Institute of Technology, Sweden}
    }

\begin{document}

\maketitle              
%

\section*{biography}


\textbf{Wenjie Liu} {received the B.Eng. degree from the University of Electronic Science and Technology of China, Chengdu, China, in 2019, and the M.Phil. degree from The Chinese University of Hong Kong, Shenzhen, in 2021. He is currently pursuing the Ph.D. degree with the School of Electrical Engineering and Computer Science, KTH Royal Institute of Technology. He was a Visiting Student/an Exchange Student at Peking University and the Harbin Institute of Technology and interned at the Intel Asia-Pacific Research and Development Center. His research interests mainly include security and privacy.}

\textbf{Panos Papadimitratos} {is a professor with the School of Electrical Engineering and Computer Science (EECS) at KTH Royal Institute of Technology, Stockholm, Sweden, where he leads the Networked Systems Security (NSS) group. He earned his Ph.D. degree from Cornell University, Ithaca, New York, in 2005. His research agenda includes a gamut of security and privacy problems, with an emphasis on wireless networks. He is an IEEE Fellow, an ACM Distinguished Member, and a Fellow of the Young Academy of Europe.}

\section*{Abstract}
\Acpl{gnss} are indispensable for various applications, but they are vulnerable to spoofing attacks. The original \ac{raim} was not designed for securing \ac{gnss}. In this context, \ac{raim} was extended with wireless signals, termed \acpl{sop}, or onboard sensors, typically assumed benign. However, attackers might also manipulate wireless networks, raising the need for a solution that considers untrustworthy \acpl{sop}. To address this, we extend \ac{raim} by incorporating all \emph{opportunistic information}, i.e., measurements from terrestrial infrastructures and onboard sensors, culminating in one function for robust \ac{gnss} spoofing detection. The objective is to assess the likelihood of \ac{gnss} spoofing by analyzing locations derived from extended \ac{raim} solutions, which include location solutions from \ac{gnss} pseudorange subsets and wireless signal subsets of untrusted networks. Our method comprises two pivotal components: subset generation and location fusion. Subsets of ranging information are created and processed through positioning algorithms, producing temporary locations. Onboard sensors provide speed, acceleration, and attitude data, aiding in location filtering based on motion constraints. The filtered locations, modeled with uncertainty, are fused into a composite likelihood function normalized for \ac{gnss} spoofing detection. Theoretical assessments of \ac{gnss}-only and multi-infrastructure scenarios under uncoordinated and coordinated attacks are conducted. The detection of these attacks is feasible when the number of benign subsets exceeds a specific threshold. A real-world dataset from the Kista Science City area is used for experimental validation. Comparative analysis against baseline methods shows a significant improvement in detection accuracy achieved by our Gaussian Mixture \ac{raim} approach. Moreover, we discuss leveraging \ac{raim} results for plausible location recovery. The theoretical analysis and experimental validation underscore the efficacy of our spoofing detection approach.

\glsresetall

\section{Introduction}
Authentication mechanisms \cite{Gmv:J21a,Gmv:J21b} improve the security of next-generation \ac{gnss}, making it more challenging for attackers to manipulate the location and time of victims \cite{Goo:J22,Wer:J22,SpaPap:C23}. However, it is crucial to recognize that authentication has limitations against sophisticated spoofing techniques, such as relay and replay attacks occurring at both signal and message levels. They can not only record signals at one location and replay them at another location later, but also make the forged signal appear to arrive earlier than the actual signal (e.g., distance-decreasing attacks) \cite{ZhaPap:C19b,LenSpaPap:C22,ZhaLarPap:J22}. Then, attackers control one or more satellites of their choice and gradually induce location deviation on the victim receiver \cite{SheWonCheChe:C20,GaoLi:J22}. In essence, the focus should shift towards mitigating evolving threats, including slow-varying relay and replay attacks and potential exploits at the physical layer of wireless signals \cite{WesRotHum:J12,ZhaTuhPap:C15,AndCarDevGil:C17,LiuCheYanShu:C21} and overall weed out signals manipulated by the adversary. 

\Ac{raim} \cite{Bro:J92} is originally an approach to assess the integrity of \ac{gnss} signals, but it is also commonly used to detect and possibly exclude one or more satellite signals that are faulty or possibly spoofed \cite{ZhaTuhPap:C15,ZhaPap:J21}. \ac{raim} compares observed pseudoranges with expected values, and based on redundant satellite combinations, it generates the corresponding \ac{gnss}-computed locations. It relies solely on the \ac{gnss} infrastructure, offering the advantage of independence from other infrastructures, especially when leveraging multiple constellations \cite{ZhaPap:J21}. However, its inherent limitation becomes apparent when facing strong adversary attacks, targeting a significant subset of satellite signals: \ac{raim} \cite{Bro:J92} needs a minimum of 5 available (unaffected by the attack) satellites to detect a single fault or at least 6 satellites to isolate it. 

On the other hand, \ac{gnss} receivers are commonly integrated into mobile computing platforms equipped with diverse network interfaces and onboard sensors \cite{SheWonCheChe:C20,KasKhaAbdLee:J22,OliSciIbrDip:J22}. These platforms not only receive signals from satellites but also capture \acpl{sop} from cellular network \acpl{bs}, wireless \acpl{ap}, and other terrestrial networking infrastructures. By leveraging redundant sensor information and \acpl{sop}, i.e., \emph{opportunistic information}, it is possible to extend \ac{raim} mechanisms \cite{KhaRosLanCha:C14,MaaKas:J21}. \ac{raim} with inertial sensor coupling \cite{KhaRosLanCha:C14} introduced a test statistic utilizing a weighted norm of a residual vector, representing the difference between actual \ac{gps} measurements and predictions, with a threshold determining the hypothesis test. \Ac{sop}-based \ac{raim} \cite{MaaKas:J21} utilizes cellular networks as sources of \acpl{sop} and an \ac{imu}, offering a relatively accurate backup location when \ac{gnss} signals are unreliable.

The aforementioned methods (including our previous work \cite{LiuPap:C23}) assume that \acpl{sop} are benign. Onboard sensors can be trusted as long as the \ac{gnss} receiver-bearing platform is not hacked, and it is reasonable to assume in certain settings that \acpl{sop} are not adversarial. However, the challenge arises when parts of the terrestrial infrastructures are not benign or their signaling/messaging is under attack. For example, attackers can break the authentication protocols of wireless networks, such as WPA2 \cite{VanPie:C17}, replay signals, or deploy rogue 4G/5G/Wi-Fi stations \cite{ShaBorParSei:C18,SaeMooPerSho:J20,YanYanYanSon:J22}. To the best of our knowledge, the detection of \ac{gnss} attacks using untrusted \emph{opportunistic information} remains unsolved. Therefore, our work extends \ac{raim} by integrating all \emph{opportunistic information}, including terrestrial wireless signals and onboard sensors, while considering the possibility of untrustworthy wireless signals. 

In this paper, we propose and investigate an extended \ac{raim}-based \ac{gnss} spoofing detection algorithm that combines signals from \ac{gnss} satellites, available terrestrial networks (e.g., Wi-Fi and cellular), and onboard inertial sensors. The key idea is to consider different combinations of \ac{gnss} pseudoranges and ranging information extracted from untrusted network infrastructures and derive multiple location solutions with uncertainty in a \ac{raim} manner. These solutions are then aggregated into a Gaussian mixture function to evaluate the correctness of the \ac{gnss} location. It is important to note that apart from spoofing the \ac{gnss} location, we assume the attacker has rogue Wi-Fi and cellular stations, or replays legitimate terrestrial wireless network traffic. 

Our scheme consists of two key components: subset generation for temporary position computations and location fusion for spoofing detection. The first component of our scheme collects ranging information in real-time from multiple infrastructures including \ac{gnss} satellites and terrestrial infrastructures. For individual infrastructures, subsets are generated as combinations of ranging information, with a subset size from the minimal measurements required by the positioning algorithm to the maximum. Then, a positioning algorithm (multilateration, nonlinear least squares, etc.) computes temporary locations for each subset and infrastructure individually. The second component uses onboard sensors to collect speed, acceleration, and attitude data, which is used to assist in filtering the temporary locations obtained from the first component. The filtering process smoothens the locations according to motion constraints, and the uncertainty of the filtered locations is modeled as variance based on the dilution of precision or the residuals between the historical temporary locations and filtered locations. The filtered locations are then fused with uncertainty, and the composite likelihood function is normalized to obtain the ultimate likelihood function used for \ac{gnss} spoofing detection. 

Within our experiments and analysis, we initially conduct a theoretical assessment of \ac{gnss}-only and multi-infrastructure scenarios under two types of attacks: uncoordinated attacks and coordinated attacks. Coordinated attacks involve the coordinated manipulation of multiple pseudoranges and other ranging information for a specifically designed spoofing path, presenting a higher level of complexity, and uncoordinated attacks otherwise. Our analysis reveals that the proposed \ac{raim} extension generates a significantly larger set of subsets for cross-validating locations, effectively countering attacks; thus raising the bar for the attacker needing to intrude or attack different (all) types of terrestrial infrastructures and \ac{gnss}. We evaluate our method using a real-world dataset, collected at the Kista Science City, encompassing ground truth location traces, pseudoranges, acceleration, attitude, cellular names, Wi-Fi names, and \ac{rssi} measurements. Through comparative analysis with state-of-the-art baseline methods, we find improved detection accuracy achieved by our method.

In conclusion, the novelty is to handle untrusted \emph{opportunistic information} and propose a Gaussian mixture \ac{raim} for detecting \ac{gnss} spoofing attacks. We discuss how to use the results from \ac{raim} towards recovering a plausible position, in Sec.~\ref{prosch}. Additionally, we provide a theoretical analysis of the detection in Sec.~\ref{theana} and demonstrate the performance of our spoofing detection approach in Sec.~\ref{experi}. 

\section{Related Work}
\subsection{GNSS Spoofing vs Meaconing}
\Ac{gnss} spoofing generates false but correctly formatted \ac{gnss} signals \cite{HumLedPsiOha:C08,SatStrLenRan:C22}. It is difficult to detect as it can usually cause subtle discrepancies in timing, strength, and direction which most modern receivers can not distinguish. Despite the implementation of \ac{gnss} authentication measures \cite{Gmv:J21a,Gmv:J21b} for countering spoofing, attackers employ replaying techniques to mimic genuine \ac{gnss} signals \cite{MaiFraBluEis:C18,LenSpaPap:C22}, or distance-decreasing attack \cite{ZhaPap:C19b,ZhaLarPap:J22} to alter signals to create the illusion of an earlier arrival compared to their actual arrival. \citeA{SheWonCheChe:C20,GaoLi:J22} focus on a slowly varying algorithm to evade tightly-coupled \ac{gnss}/\ac{imu} systems, while \citeA{GaoLi:J23} explores two time-based gradual spoofing algorithms designed for \ac{gnss} clock in the navigation message. Moreover, a significant concern lies in the adaptability of attackers utilizing versatile \acpl{sdr}. \Acpl{sdr} not only enable \ac{gnss} spoofing, but also pose a threat to other \ac{sop} navigation, including Wi-Fi and cellular networks  \cite{ShiDavSag:J20}. This multidimensional threat underscores the need for comprehensive countermeasures.
\subsection{RAIM Protecting GNSS}
\Ac{raim} traditionally relies on redundant information and consistency checks among pseudoranges or position solutions obtained from subsets of visible satellites. One significant approach employs statistical hypothesis testing techniques, where residual errors from the positioning least-squares method are scrutinized to identify faulty measurements, and then testing is applied to assess the reliability of these measurements \cite{JoeChaPer:J14}. Clustering \ac{raim}, as proposed in \citeA{ZhaTuhPap:C15}, integrates \ac{raim} with a clustering algorithm to achieve a binary classification of location solutions from various pseudorange subsets. Their follow-up work \cite{ZhaPap:C19} digs into multi-constellation situations and \cite{ZhaPap:J21} proposes a faster fault exclusion algorithm considering performance trade-off. \Ac{ekf}-combined \ac{raim} in \citeA{KhaRosLanCha:C14,RoyFar:C17} detects and eliminates outliers in \ac{gps} and inertial sensors navigation systems using a sliding window filter. To extend \ac{raim} with more constellations, advanced \ac{raim} \cite{BlaWalEngLee:J15} is proposed to use \ac{gps}, Galileo, and others for fault exclusion. Recently, \ac{sop} techniques have also been explored. The work \cite{MaaKas:J21} focuses on \ac{raim} using cellular signals and proposes a \ac{ekf}-based navigation framework that incorporates the vehicle's kinematics model, clock errors, and cellular pseudorange measurements. None of these methods account for \ac{raim} using untrusted opportunistic information. 

\section{System Model and Adversary}
\begin{figure}
\begin{centering}
\includegraphics[width=.75\columnwidth]{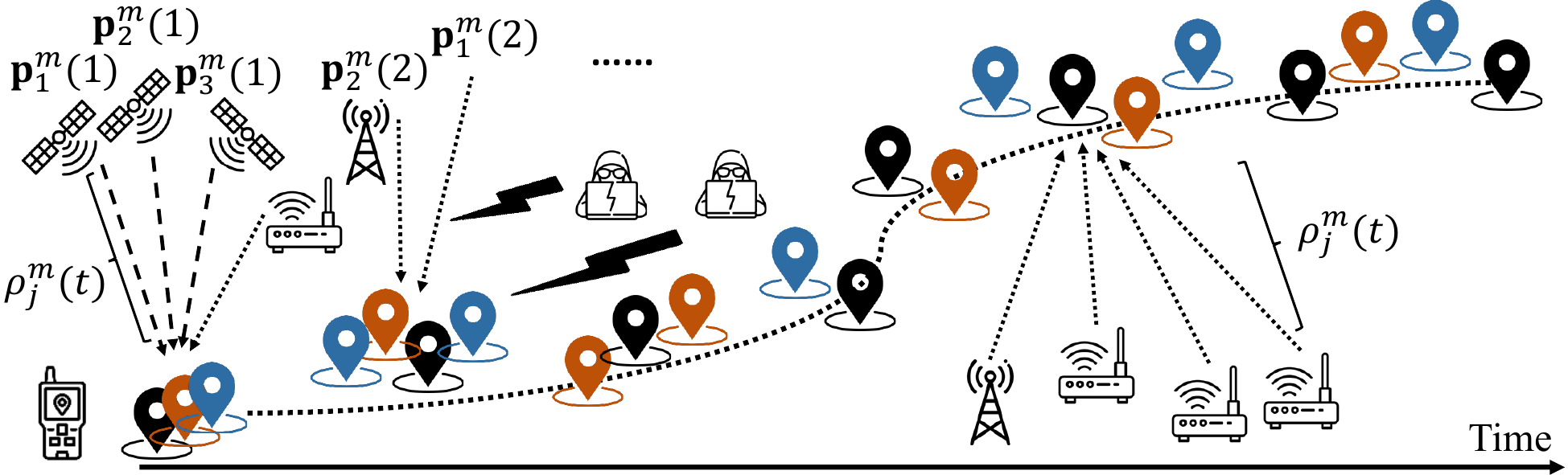}
\par\end{centering}
\caption{System and adversary model illustration.}
\label{fig:advers}
\end{figure}

\subsection{System Model}
We examine a mobile \ac{gnss} platform depicted in Fig.~\ref{fig:advers}, which can be a device such as a smartphone, car, or drone. The platform is equipped with modules that provide opportunistic information, including wireless signals from network interfaces, such as Wi-Fi and cellular networks, and motion measurements from onboard sensors, such as \acpl{imu} and wheel speed sensors. However, in some cases, \ac{gnss} satellites and network data may be unavailable, under attack, or have an unacceptable latency, leading to the provision of incorrect ranging information (e.g., pseudoranges and \ac{rssi}). On the other hand, the \ac{imu} provides three-axis acceleration measurements at a high frequency, and wheel speed sensors or pedometers provide speed measurements in certain situations, such as in vehicles or smartphones.

It is worth noting that opportunistic information is not specifically designed for localization, and is unintendedly available. In benign environments where the mobile platform is moving and navigating on a path, the estimated \ac{gnss} location would be consistent with the opportunistic information. However, when an adversary launches an attack, the \ac{gnss}-derived results could deviate from the actual location and be inconsistent with opportunistic information.

$\mathbf{p}_{\text{c}}(t) \in \mathbb{R}^3$ denotes the receiver actual location at time $t$, to be determined through positioning. \ac{gnss} and network interfaces, namely Wi-Fi and cellular, can provide ranging information from either \ac{gnss} or terrestrial networks (pseudoranges or \acpl{sop}): $\rho_j^m(t),j=1,2,...,J_m,m=1,2,...,M$ is ranging information at time $t$, where $J_m$ is the total number of \emph{anchors} from the $m$th infrastructure (i.e., \ac{gnss} satellites/\acpl{bs}/\acpl{ap}) providing ranging information and $M$ is the total number of infrastructures; $t=1,2,...,N$ and $N$ is the last time index. Correspondingly, $\mathbf{p}_j^m(t) \in \mathbb{R}^3$ is the location of the anchor for ranging information $\rho_j^m(t)$, which is known. Meanwhile, onboard sensors provide motion measurements, i.e., speed $\mathbf{v}(t)$ and acceleration $\mathbf{a}(t)$. They are updated as discrete time series at different frequencies.

\subsection{Adversary}
The adversary in Fig.~\ref{fig:advers} spoofs \ac{gnss} and network signals to force the mobile platform to get incorrect ranging information, in order to manipulate it into estimating an incorrect location. The adversary also possesses knowledge of the victim's actual location with sufficient low observational error. To carry out the attack, the adversary uses \ac{sdr} devices with state-of-the-art \ac{gnss} and network infrastructure spoofing algorithms, to falsify some wireless signals of certain anchors (i.e., \ac{gnss} satellites, \acpl{bs}, and \acpl{ap}) \cite{ShiDavSag:J20,LenSpaPap:C22}, causing a designed gradual deviation from the actual location. 

\textbf{Assumptions.} We assume the attacker is a sophisticated \ac{gnss} spoofer, rather than an unskillful jammer, with the capability to track satellites and generate multiple adversarial signals. It can also deploy rogue \acpl{ap} or \acpl{bs}. The terrestrial infrastructures are assumed to implement standardized and regularly updated security mechanisms, including authentication. Consequently, we assume the attacker uses replay/relay-like attacks \cite{ShiDavSag:J20}. Additionally, we assume that the attacker lacks physical control over the victim, ensuring the security of the procedure for deriving locations from pseudoranges and opportunistic information. 

\textbf{Attack strategy.} 
The attacker uses replay/relay-like attacks and rogue \acpl{ap}/\acpl{bs} for \ac{gnss} and network infrastructures and designs a trajectory for the spoofed locations that carefully mimics part of the wireless signals for the mobile platform to cause subtle yet significant deviations. The attacker will control part of the anchors \cite{RanOlaCap:C16,GaoLi:J22} and use various trajectory strategies, such as gradual deviating \cite{SheWonCheChe:C20} and path drift \cite{NarRanNou:C19}, to hinder the victim's ability to detect the attack on \ac{gnss} signals. These tactics aim to manipulate the mobile platform's position estimation and deceive the victim into believing that they are following the attacker-designed path (as the result of the attack).

\section{Problem Statement}
Our goal is to analyze whether all computed \ac{raim} positions are consistent with each other through using opportunistic information and accordingly make a decision on whether the current \ac{gnss} location is the result of a spoofing attack. We care most about safeguarding the \ac{gnss} position because it is usually the most accurate location provider. We would like to provide a likelihood estimate of whether the \ac{gnss} system is under attack and maximize the accuracy of \ac{gnss} spoofing detection. We focus on the design of the extended \ac{raim}. 

At a time $t$, we use data $\{\mathbf{p}_j^m(i), \rho_j^m(i), {\mathbf{v}}(i), {\mathbf{a}}(i)\}$ for $j=1,2,...,J_m,m=1,2,...,M$ and $0<i<t$, to determine whether the current \ac{gnss} position is affected by an attack. We formulate two hypotheses:
\begin{itemize}
    \item $H_0$: \ac{gnss} is not under attack.
    \item $H_1$: \ac{gnss} is under attack.
\end{itemize}
The decision at time $t$ is denoted as $\hat{H}(t)\in \{H_0,H_1\}$. We define true positive for $t$ as $\mathbb{I}\{\hat{H}(t)=H_1|H_1\}=1$ and false positive as $\mathbb{I}\{\hat{H}(t)=H_1|H_0\}=1$, where the indicator function $\mathbb{I}\{A|B\}$ takes value 1 if A holds on condition of B. We denote the total number of positives as $N_\text{P}$, the number of true positives as $N_\text{TP}$, and the number of false positives as $N_\text{FP}$. Then, the true positive probability for the period $0<t \le N$ is
$P_\text{TP} (\hat{H}(t)) = P(\hat{H}(t)=H_1|H_1)=\frac{N_\text{TP}}{N_\text{P}}$
and the Type I error (false positive) probability is
$P_\text{FP} (\hat{H}(t)) = P(\hat{H}(t)=H_1|H_0)=\frac{N_\text{FP}}{N-N_\text{P}}$
The objective is to maximize the true positive probability $P_\text{TP}$ of attack detection, given $P_{\text{FP}_{\max}}$:
$$
\begin{matrix}
	\text{max}&P_\text{TP} (\hat{H}(t))\\
	\text{s.t.}&P_\text{FP} (\hat{H}(t)) \le P_{\text{FP}_{\max}}\\
\end{matrix}.
$$

\section{Proposed Scheme}
\label{prosch}

\subsection{Scheme Outline}
\label{schout}
\begin{figure}
\begin{centering}
\includegraphics[width=.9\columnwidth]{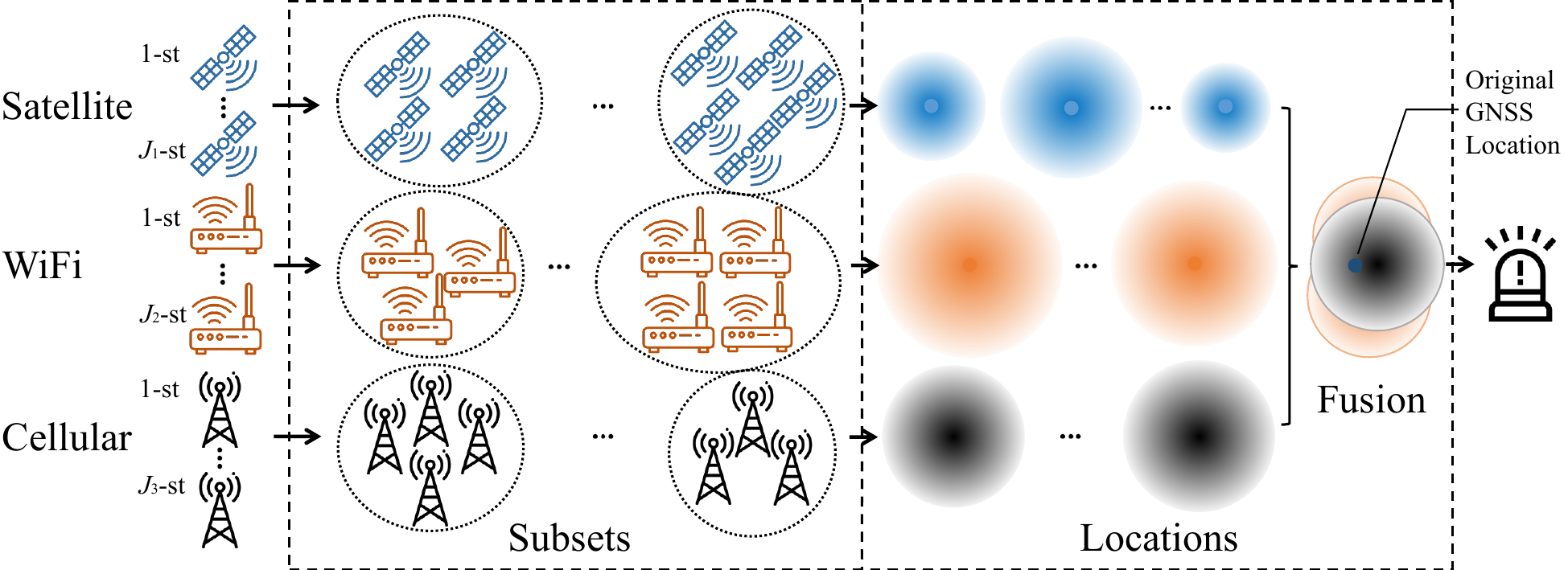}
\par\end{centering}
\caption{System overview of Gaussian mixture \ac{raim}.}
\label{fig:scheme}
\end{figure}
We propose a Gaussian mixture \ac{raim} approach for detecting \ac{gnss} spoofing by utilizing multiple information sources, such as \ac{gnss} pseudoranges, \acpl{sop}, speed, acceleration, and attitude. Fig.~\ref{fig:scheme} provides a high-level overview of our \ac{gnss} spoofing detection system, which collects input data from \ac{gnss} and other opportunistic information sources. Creating subsets of ranging information based on different infrastructures allows us to accommodate variations in the data characteristics of each information source, including considerations of average distance, variance, and accuracy of distance. Utilizing the corresponding positioning algorithm, we compute temporary locations for each subset, which are then ``de-noise'' using an onboard sensor-assisted filter and modeled with a statistical tool to estimate their uncertainty. By comparing the fused locations with uncertainty to the original \ac{gnss} computed location, we can assess the likelihood of \ac{gnss} spoofing.

Our scheme consists of two key components: subset generation for positioning and location fusion for spoofing detection. The overall procedure is shown as Algorithm \ref{alg:extraim}.

\begin{algorithm}
\hspace*{\algorithmicindent} \textbf{Input} $\{\mathbf{p}_j^m(t), \rho_j^m(t), {\mathbf{v}}(t), {\mathbf{a}}(t)\}$ for $j=1,2,...,J_m$ and $m=1,2,...,M$\\
\hspace*{\algorithmicindent} \textbf{Parameter} $\Lambda$\\
\hspace*{\algorithmicindent} \textbf{Output} \textit{IsAttack}
\begin{algorithmic}[1]
\For{$t = 1$, $t{+}{+}$, $t \le N$} \Comment{Time index}
\For{$l = 1$, $l{+}{+}$, $l \le L$} \Comment{Loop through all subsets}
\State $\mathbf{p}_l(t) \gets$ Eq.~\eqref{eq:tdoa}, \eqref{eq:multil}, or \eqref{eq:geoloc} \Comment{Positioning algorithms}
\State $\hat{\mathbf{p}}_l(t) = \mathbf{W}\mathbf{t} \gets$  Eq.~\eqref{eq:proall} \Comment{Local polynomial regression}
\State $\boldsymbol{\sigma}_l (t) \gets$ Eq.~\eqref{eq:sigmalt} \Comment{Uncertainty modelling}
\State $f_{l,t}(\mathbf{p})={\frac {1}{\boldsymbol{\sigma}_l (t) {\sqrt {2\pi }}}}\exp \left(-{\frac {1}{2}}{\frac {(\mathbf{p} - \hat{\mathbf{p}}_l(t) )^{2}}{\boldsymbol{\sigma}_l (t) ^{2}}}\right)$ \Comment{Probability density function}
\EndFor
\State $f_t(\mathbf{p})=\frac{1}{L} \sum_{l=1}^L f_{l,t}(\mathbf{p})$ \Comment{Gaussian mixture function}
\If{$f_t(\hat{\mathbf{p}}_{\text{c}}) < \Lambda$} 
    \State \textit{IsAttack} $=$ \textit{True}
\Else
    \State \textit{IsAttack} $=$ \textit{False}
\EndIf 
\EndFor
\end{algorithmic}
\caption{Extending RAIM with a Gaussian mixture of
opportunistic information \label{alg:extraim}}
\end{algorithm}

The first component (Sec.~\ref{subsec:subgen}) of our scheme collects ranging information in real-time from multiple infrastructures, including \ac{gnss}, Wi-Fi, and cellular networks. For different infrastructures, subsets are generated as combinations of ranging information. Note that \ac{gnss} includes \ac{gps}, Galileo, and so on, so we can also see them as different infrastructures. The size of the subsets is from the minimal measurements required by the positioning algorithm to the maximum. Then, it computes temporary locations with uncertainty for each subset and infrastructure individually. 

The second component (Sec.~\ref{subsec:locfus}) utilizes onboard sensors to collect speed, acceleration, and attitude data, which is used to assist in the refinement of temporary locations obtained from the first component. The filtering procedure smooths the locations according to the motion constraints of the receiver. The filtered locations are then fused with the resultant positioning uncertainty, and the composite likelihood function is normalized to derive the final likelihood function utilized for \ac{gnss} spoofing detection. 

\subsection{Subset Generation}
\label{subsec:subgen}
The first component, subset generation, is actually a resampling method that selects different subsets of all the ranging information, similar to cross-validation. 
\subsubsection{Raw Data Collection}
The input raw data are pseudoranges, \acpl{sop} (providing distance estimates of the device from the corresponding terrestrial infrastructure elements (anchors)), and motion measurements. Pseudoranges are not actual distances between satellites and the \ac{gnss} receivers and are often estimated using the time the navigation signals take to propagate in the sky, which have random and systematic clock errors. \acpl{sop} from Wi-Fi and cellular anchors contain \ac{rssi}, \ac{snr}, \ac{toa}, and radio type. These values, even though they are not actual distances, are related to distances, forming the fundamental basis for positioning algorithms: 
\begin{itemize}
    \item Pseudorange is an approximation of the distance measurement between a satellite and a \ac{gnss} receiver. The pseudorange for each satellite is the time (the signal takes to reach the receiver) multiplied by the speed of light: $\rho=c\Delta t$. However, because the quartz oscillator of the receiver clock is not accurate enough, the error may be hundreds of meters after one second, i.e., not the actual distance. Fortunately, the clock is used to measure the pseudoranges at the same time, so all pseudoranges have the same clock error. By finding the pseudoranges of at least four anchors, the clock error can be estimated. 
    \item \acpl{sop} can be used for both direct and indirect positioning algorithms, such as Geolocation \cite{Mozilla2023}, fingerprint-based, and range-based localizations. Geolocation localization directly uses \ac{rssi} (as well as \ac{snr} and signal frequency if available), based on which weighted nonlinear least squares estimate the location of the receiver. Range-based localization \cite{LiuChe:J21} needs the approximations of the distances between \acpl{bs}/\acpl{ap} and the receiver. \ac{rssi} can estimate the distance between a \ac{bs}/\ac{ap} and the receiver based on the \ac{rssi} and the path loss model. The path loss model accounts for large-scale fading and random noise. The large-scale fading is stable and can be measured in advance in an environment. Random noise can be reduced by multiple measurements and least squares. The receiver can be located by finding the distances of at least three \acpl{bs}/\acpl{ap}.
\end{itemize}
\subsubsection{Size and Combination}
As we do not have any assumption on the number of falsified ranges, the sizes and combinations of generated subsets should explore all possibilities, from the minimum required by positioning to the maximum: For \ac{gnss}, three pseudoranges from satellites will determine the location. latitude, longitude, and height, while the fourth one will synchronize the receiver clock. For \acpl{bs}/\acpl{ap}, at least three distances can determine the receiver location and the clock error has relatively little effect. Thus, the sizes of subsets are $C(J_1, 4)+C(J_1, 5)+...+C(J_1, J_1)$ for \ac{gnss}, $C(J_m, 3)+C(J_m, 4)+...+C(J_m, J_m),m\in\{2,3\}$ for Wi-Fi and cellular networks, and the total number of subsets is denoted as $L$. Additionally, if necessary, we can further categorize the information from \ac{gps}, Galileo, 2G, 5G, etc, as distinct types.
\subsubsection{Positioning Methods}
For computing temporary locations from subsets of ranging information, various methods such as \ac{tdoa}, multilateration, and weighted nonlinear least squares are employed exclusively for their targeted data. 
\begin{itemize}
    \item \ac{tdoa} localization uses the difference of distances to locate the mobile platform, which solves a number of hyperbolic equations:
    \begin{equation}
        \rho_j^m(t)-\rho_k^m(t)=||\mathbf{p}_j^m(t)-\hat{\mathbf{p}}_{\text{c}}(t)||_2 - ||\mathbf{p}_k^m(t)-\hat{\mathbf{p}}_{\text{c}}(t)||_2 ,  \forall j, k
    \label{eq:tdoa}
    \end{equation}
    If we would like to estimate time, substitute $\rho=c\Delta t$ into the above. 
    \item Distance-based multilateration localization uses multiple distances between the mobile platform and multiple \acpl{bs}/\acpl{ap} with known locations:
    \begin{equation}
        \rho_j^m(t)=||\mathbf{p}_j^m(t)-\hat{\mathbf{p}}_{\text{c}}(t)||_2, \forall j
    \label{eq:multil}
    \end{equation}
    \item \revadd{Geolocation localization uses weighted nonlinear least squares to minimize the weighted sum of the squared distances between \acpl{bs}/\acpl{ap} and the estimated location. The weights are inversely proportional to the squared signal strengths \cite{Mozilla2023}.
    \begin{equation}
    \underset{\hat{\mathbf{p}}_{\text{c}}(t)}{\mathop{\min}}\quad \sum_{j} \left(\frac{||\mathbf{p}_j^m(t)-\hat{\mathbf{p}}_{\text{c}}(t)||_2}{\rho_j^m(t)}\right)^2
    \label{eq:geoloc}
    \end{equation}
    }
\end{itemize}

Each subset computes an estimated location of the mobile platform $\mathbf{p}_l(t)$, where $l=1,2,...,L$ and $L$ is the total number of subsets.  

\subsection{Location Fusion}
\label{subsec:locfus}
The second component, location fusion, utilizes signal processing techniques to estimate the likelihood of \ac{gnss} spoofing.
\subsubsection{Motion Information}
We use $\mathbf{p}_l(t)$ from the previous component and ${\mathbf{v}}(t)$, ${\mathbf{a}}(t)$ from onboard sensors as input data and local polynomial regression to filter the noise. 

The state of mobile platform, $\big(\mathbf{p}_l(t),\mathbf{v}(t),\mathbf{a}(t) \big),l=1,2,...,L$, evolved from $t-1$, so:
\begin{align}
\mathbf{p}_l(t)&=\mathbf{p}_l(t-1)+\mathbf{v}(t-1)+\frac{1}{2}\mathbf{a}(t-1)\\
\mathbf{v}(t)&=\mathbf{v}(t-1)+\mathbf{a}(t-1)
\end{align}
and the state transition matrix is:
\begin{equation}
    \mathbf {F}(t) ={\begin{bmatrix}\mathbf{1}&\mathbf{1}\\\mathbf{0}&\mathbf{1}\end{bmatrix}}
\end{equation}
and the control-input matrix is $\mathbf {B}(t) =\begin{bmatrix}\mathbf{\frac{1}{2}} & \mathbf{1}\end{bmatrix} ^\text{T}$. We conclude that $\big(\mathbf{p}_l(t),\mathbf{v}(t)\big)=\mathbf {F}(t-1) \cdot \big(\mathbf{p}_l(t-1),\mathbf{v}(t-1)\big)+\mathbf{B}(t-1) \cdot \mathbf{a}(t-1)+\mathbf {n} $, where $\mathbf {n}$ models noise. 

Then, for filtering the noise, we use an estimator $\hat{\mathbf{p}}_l(t) = \mathbf{W}\mathbf{t}$, where $\mathbf{W} \in \mathbb{R} ^{2 \times (n+1)}$ is a matrix of polynomial coefficients, $\mathbf{t}$ is a $(n+1)$ dimensional vector and $[\mathbf{t}]_i=t^{i-1}$, and $\mathbf{W}$ is from the local polynomial regression problem:
\begin{equation}
    \begin{array}{*{20}{c}}
    {\mathop {\min }\limits_{\mathbf{W}} }&{\sum\limits_{t=t'-w}^{t'-1} [\mathbf{W} \mathbf{t}-\mathbf{p}_l(t)]^\top K_\text{loc}(t-t')[\mathbf{W} \mathbf{t}-\mathbf{p}_l(t)]} \\ 
    {{\text{s}}{\text{.t}}{\text{.}}}&{|\mathbf{W} \mathbf{t'} - \mathbf{p}}_l(t' - 1)| \le \boldsymbol{\epsilon}
    \end{array}
\label{eq:proall}
\end{equation}
where $w$ is a rolling window of the filter and $\boldsymbol{\epsilon} \in \mathbb{R}^2$ in the constraint is a small tolerance. After solving the problem, we have $\hat{\mathbf{p}}_l(t)$.

\subsubsection{Uncertainty Modelling}
We have different cases: (i) For statistical positioning techniques, the covariance matrix from the positioning process derives confidence intervals; (ii) For other positioning techniques, we use the residual vector from the local polynomial regression in \eqref{eq:proall} to model the uncertainty of estimated locations $\hat{\mathbf{p}}_l(t)$. 

\revadd{In case (i), \ac{dop} derived from the covariance matrix of statistical positioning techniques serves as a direct indicator of the uncertainty associated with positioning results. To illustrate, the position \ac{dop} can be determined by considering the diagonal elements of the covariance matrix obtained via the least squares method. This value is often correlated with the standard deviation $\boldsymbol{\sigma}_l (t)$ of $\hat{\mathbf{p}}_l(t)$.}

In case (ii), the residual part between $\mathbf{p}_l(t)$ and $\hat{\mathbf{p}}_l(t)$ from $l$th subset at time $t$ is
\begin{equation}
    \mathbf{d}_l(t)=\hat{\mathbf{p}}_l(t)-\mathbf{p}_l(t).
\end{equation}
Then, the uncertainty is the standard deviation $\boldsymbol{\sigma}_l (t)$, 
\begin{equation}
    \left ( \boldsymbol{\sigma}_l (t) \right )^2=\frac{1}{L}\sum_{l=1}^L\left (\mathbf{d}_l(t) - \frac{1}{L}\sum_{i=1}^L\mathbf{d}_i(t) \right )^2
\label{eq:sigmalt}
\end{equation}
and thus the probability density function is $f_{l,t}(\mathbf{p})={\frac {1}{\boldsymbol{\sigma}_l (t) {\sqrt {2\pi }}}}\exp \left(-{\frac {1}{2}}{\frac {(\mathbf{p} - \hat{\mathbf{p}}_l(t) )^{2}}{\boldsymbol{\sigma}_l (t) ^{2}}}\right)$, where the operations are point-wise. 

\subsubsection{Gaussian Mixture}
The Gaussian mixture function fuses temporary locations with uncertainties derived from the subsets. So far we have locations $\hat{\mathbf{p}}_l(t)$ associated with uncertainty $\boldsymbol{\sigma}_l (t)$, which are assumed to follow distributions $\mathcal{N}(\hat{\mathbf{p}}_l(t), \boldsymbol{\sigma}_l (t))$. At time $t$, the Gaussian mixture for $l=1,2,...,L$ can be represented as: 
\begin{align*}
    f_t(\mathbf{p})&=\frac{1}{L} \sum_{l=1}^L f_{l,t}(\mathbf{p}) \\
    &=\frac{1}{L} \sum_{l=1}^L {\frac {1}{\boldsymbol{\sigma}_l (t) {\sqrt {2\pi }}}}\exp \left(-{\frac {1}{2}}{\frac {(\mathbf{p} - \hat{\mathbf{p}}_l(t) )^{2}}{\boldsymbol{\sigma}_l (t) ^{2}}}\right)
\end{align*}
which serves as a likelihood function; specifically, it represents the likelihood of the original \ac{gnss} computed position $\hat{\mathbf{p}}_{\text{c}}$ not being under attack, denoted as $f_t(\hat{\mathbf{p}}_{\text{c}})$. To make a decision, a threshold $\Lambda$ is predefined, and if $f_t(\hat{\mathbf{p}}_{\text{c}})$ is less than $\Lambda$, an alarm will be raised. If an attack is detected, the proposed scheme would take the value under the maximum likelihood of the mixture function $f_t(\mathbf{p})$ to recover the actual position: $\hat{\mathbf{p}}_{\text{c}}(t)=\arg \max_\mathbf{p} f_t(\mathbf{p})$. 

\subsection{Theoretical Analysis}
\label{theana}
To facilitate the analysis of the overall algorithm performance, we assign zero values to random errors while retaining the deviation induced by the attacker. Additionally, in our scenario, we consider that the satellite geometry and the geometry of other anchors for positioning are not poor. Furthermore, we categorize the attacks into (i) \emph{Coordinated Attack}, where the spoofed ranging information (across all available infrastructures) is collectively designed with a specific spoofing position in mind, and (ii) \emph{Uncoordinated Attack} otherwise (the spoofed ranging information is independently (possibly randomly) chosen). 

\subsubsection{GNSS-only Case}
For multilateration algorithm used by \ac{gnss} positioning, at least 4 satellites are needed to determine coordinates and local clock error, so we denote $N_{\text{min}}=4$ as the minimum satellite number. Similarly, $N_{\text{sat}}$ is the total number of satellites providing pseudoranges and $N_{\text{adv}}$ is the number of pseudoranges with deviations induced by the attacker.
\begin{lem}
    Suppose that $N_{\text{sat}}-N_{\text{adv}}>N_{\text{min}}$. Then, we can recover $\mathbf{p}_{\text{c}}$ from an Uncoordinated Attack.
\end{lem}
\begin{prop}
    Suppose that $N_{\text{sat}}-N_{\text{adv}} \ge N_{\text{min}}$. Then, we can detect an Uncoordinated Attack. 
\end{prop}
\begin{cor}
    An adversarial subset can contain at most $N_{\text{min}}-1$ benign pseudoranges without being detected.
\end{cor}
\begin{lem}
    Suppose that $\sum_{i=N_{\text{min}}}^{N_{\text{sat}}-N_{\text{adv}}}C(N_{\text{sat}}-N_{\text{adv}},i) > \sum_{i=1}^{N_{\text{adv}}}C(N_{\text{adv}},i)\sum_{j=N_{\text{min}}-i}^{N_{\text{min}}-1}C(N_{\text{min}}-1,j)$.\footnote{When $A>B$, the term $\sum_{i=A}^B \cdot$ takes value 1.} Then, we can recover $\mathbf{p}_{\text{c}}$ from a Coordinated Attack.
\end{lem}
This represents a classical Multiple Fault Detection and Exclusion. For the detailed proof process, please consult the single constellation case in \citeA{ZhaPap:J21}.
\subsubsection{Multi-infrastructure Case}
Most distance-based multilateration algorithms for network-based positioning need at least 3 anchors (i.e., \acpl{bs} and \acpl{ap}) to determine the coordinates of the receiver. We denote $N_{\text{min}}^m$ as the minimum anchor number required by the positioning for $m$th infrastructure, $N_{\text{anc}}^m$ as the total number of anchors providing ranging information, and $N_{\text{adv}}^m$ as the number of anchors affected by an attacker thus providing adversarial ranging information.
\begin{thm}
    Suppose that $\exists m$, s.t. $N_{\text{anc}}^m-N_{\text{adv}}^m > N_{\text{min}}^m$. Then, we can recover $\mathbf{p}_{\text{c}}$ from an Uncoordinated Attack.
    \label{thm:mulnon}
\end{thm}
\begin{thm}
    Suppose that $\sum_{m=1}^M \sum_{i=N_{\text{min}}^m}^{N_{\text{anc}}^m-N_{\text{adv}}^m}C(N_{\text{anc}}^m-N_{\text{adv}}^m,i) > \sum_{m=1}^M \sum_{i=1}^{N_{\text{adv}}^m}C(N_{\text{adv}}^m,i)\sum_{j=N_{\text{min}}^m-i}^{N_{\text{min}}^m-1}C(N_{\text{min}}^m-1,j)$. Then, we can recover $\mathbf{p}_{\text{c}}$ from a Coordinated Attack.
    \label{thm:mulcol}
\end{thm}
Intuitively, the conditions of the multi-infrastructure case are easier to satisfy than \ac{gnss}-only case, so the detector using multi-infrastructure is stronger. We provide a proof sketch: Leveraging the Lemmas derived from the \ac{gnss}-only scenario, we implement Multiple Fault Detection and Exclusion similar to the case of a single constellation, for each infrastructure within multi-infrastructures separately. The results from both individual subsets and infrastructures are then aggregated. Successful detection and exclusion are feasible as long as the benign subsets outnumber the attack(er) subsets. 

\section{Experiments}
\label{experi}
We study a dataset combining \ac{gnss} with untrusted opportunistic information, in which the trace is collected in the real world and the attacks are synthesized based on real data. Then, we evaluate the true positive probability of attack detection. 

\begin{figure}
    \centering
    \includegraphics[width=.85\columnwidth]{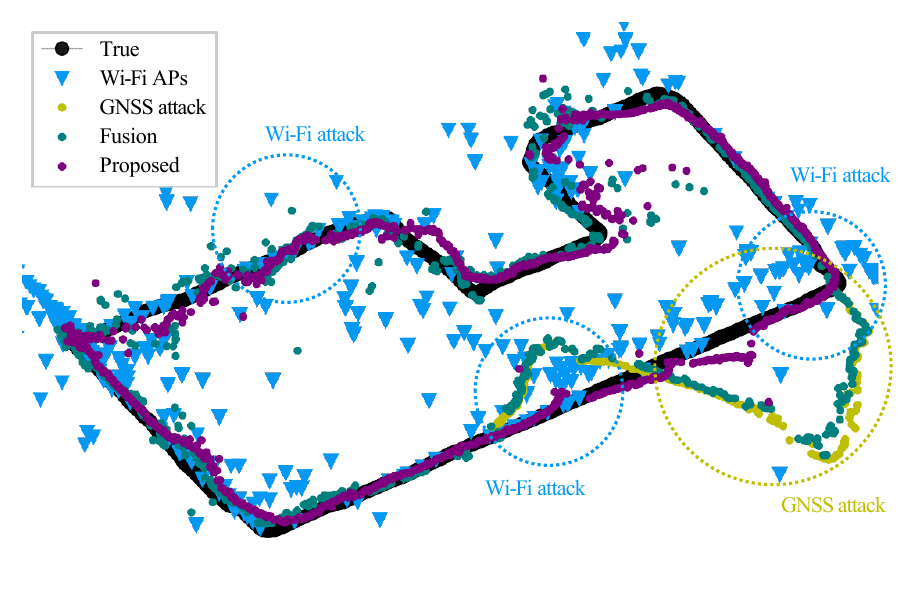}
    \caption{An illustration of the attack-induced deviations and the recovered locations from the proposed method.}
    \label{fig:traces}
\end{figure}

\subsection{Dataset}
The raw \ac{gnss} measurements are collected using Google Pixel 4 XL phone and GNSSLogger \cite{FuKhiVan:C20} in the Android system, and contain a walking trace at the Kista Science City. This dataset also incorporates data from onboard sensors, such as \ac{imu}, and wireless networks, including cellular and Wi-Fi, with details such as names, \ac{rssi}, and unique IDs. The latitude and longitude information of \acpl{ap} and \acpl{bs} is sourced from WiGLE \cite{BobArkUht:J23}. Positioning algorithms employed include weighted least squares for \ac{gnss} and weighted nonlinear least squares for network infrastructures. 

For \ac{gnss} spoofing, instead of spoofing the mobile phone, the attack is synthesized by modifying the raw \ac{gnss} data. We gradually increase the pseudorange of two satellites, causing the derived location to deviate progressively from the actual one. Additionally, three rogue Wi-Fi \acpl{ap} are synthesized and strategically ``placed'' near some locations of the trace to broadcast signals to simulate \acpl{ap} from a different fixed location, which causes 600 spoofed Wi-Fi \ac{rssi} measurements/values. The \ac{gnss} spoofing spans 130 seconds, while simulated signals from rogue Wi-Fi \acpl{ap} last around 120 seconds, with the entire trace extending over 1030 seconds. The \ac{gnss} spoofing deviation gradually increases from 0 to 145 meters and then decreases to 0 meters illustrated by Fig.~\ref{fig:traces}. 

\subsection{Baseline Methods}
To assess the performance improvement of the proposed \ac{raim}, we compare it with baseline methods:
\begin{itemize}
    \item Kalman filter: This method utilizes filtering based on \ac{imu} information in conjunction with \ac{gnss} locations. It calculates the distance between the filtered location and the original \ac{gnss} location. If the distance exceeds a specified threshold, a spoofing alarm is triggered.
    \item Location fusion: In scenarios where the mobile platform cannot obtain detailed ranging information like pseudoranges from the \ac{gnss} receiver, and thus cannot generate pseudorange subsets for \ac{raim}, this method directly acquires \ac{gnss}-provided locations and location estimations from network infrastructures. Subsequently, a fusion method \cite{LiuPap:C23} is employed to calculate an ultimate location with uncertainty for detection.
\end{itemize}

\begin{figure}
    \centering
    \begin{tikzpicture}
    \begin{axis}[
        xlabel={False positive $P_{\text{FP}_{\max}}$ [\%]},
        ylabel={True positive $P_\text{TP}$ [\%]},
        xmin=1, xmax=5,
        ymin=0, ymax=100,
        xtick={1,2,3,4,5},
        ytick={0,20,40,60,80,100},
        legend cell align={left},
        legend pos=south east,
        legend columns=1, 
        xmajorgrids=true,
        ymajorgrids=true,
        grid style=dashed,
    ]
    \addplot[
        color=mycolor1,
        mark=square,
        ]
        coordinates {
        (1,59)(2,66)(3,78)(4,88)(5,91)
        };
        \addlegendentry{Proposed with exclusion}
    \addplot[
        color=mycolor2,
        mark=triangle,
        ]
        coordinates {
        (1,52)(2,60)(3,66)(4,72)(5,80)
        };
        \addlegendentry{Proposed without exclusion}
    \addplot[
        color=mycolor3,
        mark=x,
        ]
        coordinates {
        (1,1)(2,1)(3,1)(4,1)(5,1)
        };
        \addlegendentry{Kalman filter}
    \addplot[
        color=mycolor4,
        mark=o,
        ]
        coordinates {
        (1,42)(2,52)(3,59)(4,65)(5,72)
        };
        \addlegendentry{Location fusion}
    \end{axis}
    \end{tikzpicture}
    \caption{$P_\text{TP}$ of the proposed method (with/without exclusion), Kalman filter, and location fusion-based detector.}
    \label{fig:result}
\end{figure}
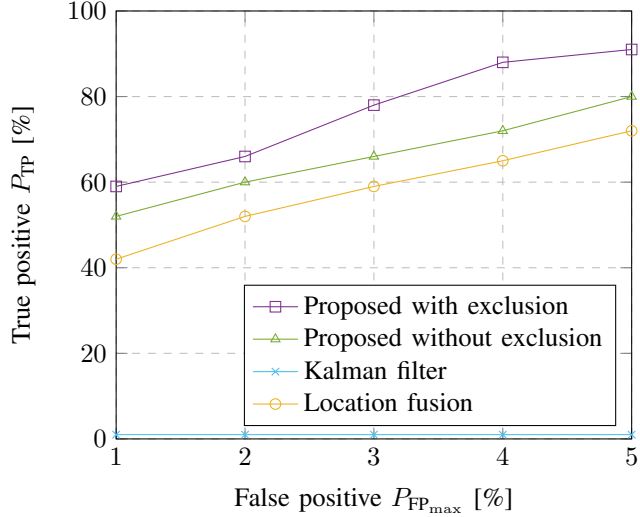

\begin{figure}
    \centering
    \includegraphics[width=.66\columnwidth]{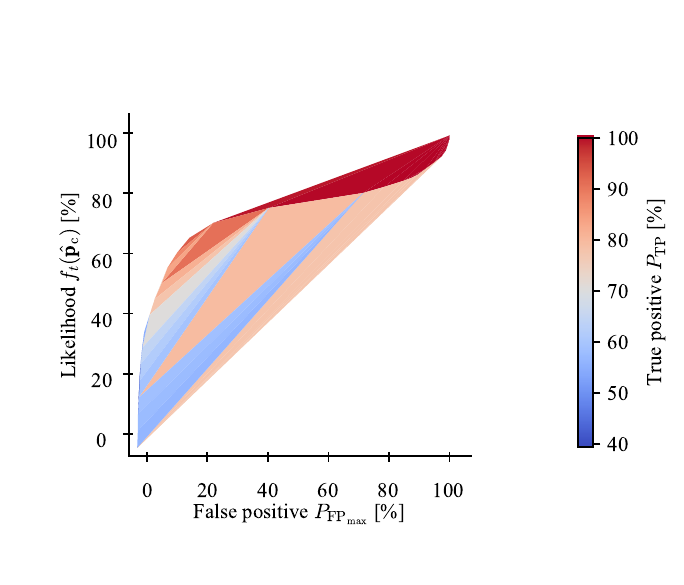}
    \caption{A plot of the relation between likelihood, $f_t(\hat{\mathbf{p}}_{\mathrm{c}})$, false positive probability, $P_{\mathrm{FP}_{\max}}$, and true positive probability, $P_\mathrm{TP}$.}
    \label{fig:likeli}
\end{figure}

\subsection{Detection Result}
Recall that our problem is maximizing the true positive probability of spoofing detection, $P_\text{TP}$, when fixing false positive, $P_{\text{FP}_{\max}}$. We choose different $P_{\text{FP}_{\max}}$ and evaluate $P_\text{TP}$. We also test two different strategies to construct the Gaussian mixture function, i.e., with and without outlier subset exclusion. If the distance between the temporary location $\hat{\mathbf{p}}_l(t),l=1,2,...,L$ and $\hat{\mathbf{p}}_{\text{c}}(t)$ is larger than a threshold value, then the exclusion strategy will drop the temporary location $\hat{\mathbf{p}}_l(t)$. 

In the presence of an attacker-designed \ac{gnss} trace, as in Fig.~\ref{fig:traces}, and attacker-replayed \acpl{sop} in the dataset, our proposed scheme exhibits a detection performance improvement of over 60\% compared to the Kalman filter as depicted in Fig.~\ref{fig:result}. One key factor contributing to this enhancement is our algorithm's integration of both onboard sensors and terrestrial network infrastructures, whereas the Kalman filter incorporates only \ac{imu} data. When compared to the location fusion approach, our proposed scheme with exclusion exhibits a 10--15\% increase in detection performance and a 5--10\% gain for the proposed without exclusion. Based on our theoretical analysis in Sec.~\ref{theana}, this improvement is attributed to our algorithm's utilization of detailed ranging information instead of relying solely on direct location data. Notably, our approach distinguishes itself by incorporating heterogeneous data, checking for inconsistencies, and avoiding the assumption that redundant wireless signals are inherently benign as references.

In addition, through an internal comparison between the proposed method with outlier subset exclusion and the proposed method without outlier subset exclusion, we observe a performance gain of 5--10\% attributable to the outlier exclusion strategy in our \ac{raim}. This gain is achieved by automatically eliminating subsets with temporary locations that are more than a threshold distance (around 150 meters in our setting) away from the peak of the Gaussian mixture function. This exclusion effectively minimizes the negative impact of fault subsets on the construction of the Gaussian mixture function. 

Fig.~\ref{fig:traces} shows the proposed recovered location, i.e., the peak of the Gaussian mixture function, from the proposed method, in which we can observe that it is resisted to gradual deviation attack in the presence of rogue \acpl{ap}/\acpl{bs}. Fig.~\ref{fig:likeli} is the relation between likelihood, $f_t(\hat{\mathbf{p}}_{\mathrm{c}})$, false positive probability, $P_{\mathrm{FP}_{\max}}$, and true positive probability, $P_\mathrm{TP}$. We notice that by setting the likelihood threshold to a relatively high value, $P_\mathrm{TP}$ can be very high while maintaining a low $P_{\mathrm{FP}_{\max}}$; which implies that the alarm should be triggered only if the likelihood of spoofing is high. 

\section{Conclusion}
The paper proposed a Gaussian mixture \ac{raim} for detecting \ac{gnss} spoofing by using multiple sources of untrusted information such as ranging information, speed, and acceleration. The system collects data from \ac{gnss}, Wi-Fi, cellular networks, and onboard sensors, generates subsets of ranging information for each infrastructure, computes temporary locations for each subset using a positioning algorithm, filters the locations using onboard sensors, models their uncertainty with statistics, and fuses them to assess the likelihood of \ac{gnss} spoofing. The approach consists of two key components: subset generation for positioning and location fusion for spoofing detection. The simulation and numerical results show a more than 90\% true positive detection rate under 5\% false alarm. 

In the continuation of this work, our attention will be directed towards enhancing accuracy, efficiency, and conducting further experiments. To elevate location and detection accuracy, we will introduce various strategies for excluding abnormal information. In terms of computational efficiency, we plan to propose sampling methods for subsets and provide theoretical evidence of their performance equivalence with/without sampling. Additionally, we aim to refine our theoretical analysis and conduct additional experiments in both simulators and real-world environments. 

\renewcommand{\APACrefbtitle}[2]{\Bem{#1}}
\bibliographystyle{apacite}
\bibliography{reference/references}

\end{document}

%% file: reference/acronym.tex
\newac{speb}{SPEB}{square position error bound}
\newac[plural=EFIMs,firstplural=Fisher information matrices (EFIMs)]{efim}{EFIM}{Fisher information matrix}
\newac{ne}{NE}{Nash equilibrium}
\newac{mse}{MSE}{mean squared error}
\newac{toa}{TOA}{time-of-arrival}
\newac{snr}{SNR}{signal-to-noise ratio}
\newac{lan}{LAN}{local area network}
\newac{psd}{PSD}{positive semidefinite}
\newac{pd}{PD}{positive definite}
\newac{wrt}{w.r.t.}{with respect to}
\newac{lhs}{L.H.S.}{left hand side}
\newac{wp1}{w.p.1}{with probability 1}
\newac{kkt}{KKT}{Karush-Kuhn-Tucker}
\newac{wlog}{w.l.o.g.}{without loss of generality}
\newac{mle}{MLE}{maximum likelihood estimation}
\newac{gps}{GPS}{global positioning system}
\newac{rssi}{RSSI}{received signal strength indicator}
\newac{mimo}{MIMO}{multiple-input multiple-output}
\newac{csi}{CSI}{channel state information}
\newac{fdd}{FDD}{frequency division duplexing}
\newac{ms}{MS}{mobile station}
\newac{bs}{BS}{base station}
\newac{d2d}{D2D}{device-to-device}
\newac{slnr}{SLNR}{signal-to-interference-leakage-and-noise-ratio}
\newac{ula}{ULA}{uniform linear antenna array}
\newac{pas}{PAS}{power angular spectrum}
\newac{mmse}{MMSE}{minimum mean square error}
\newac{zf}{ZF}{zero-forcing}
\newac{rzf}{RZF}{regularized zero-forcing}
\newac{as}{AS}{angular spread}
\newac{aod}{AOD}{angle of departure}
\newac{iid}{i.i.d.}{independent and identically distributed} 
\newac{sinr}{SINR}{signal-to-interference-and-noise ratio}
\newac{tdd}{TDD}{time-division duplex}
\newac{rvq}{RVQ}{random vector quantization}
\newac{rhs}{R.H.S.}{right hand side}
\newac{mrc}{MRC}{maximum ratio combining}
\newac{cdf}{CDF}{cumulative distribution function}
\newac{a.s.}{a.s.}{almost surely}
\newac{los}{LOS}{line-of-sight}
\newac{jsdm}{JSDM}{joint spatial division and multiplexing}
\newac{map}{MAP}{maximum a posteriori}
\newac{klt}{KLT}{Karhunen-Lo\`eve Transform}
\newac{lbe}{LBE}{link bargaining equilibrium}
\newac{se}{SE}{Stackelberg equilibrium}
\newac{uav}{UAV}{unmanned aerial vehicle}
\newac{nlos}{NLOS}{non-line-of-sight}
\newac{pdf}{PDF}{probability density function}
\newac{em}{EM}{expectation-maximization}
\newac{knn}{KNN}{$k$-nearest neighbor}
\newac{svd}{SVD}{singular value decomposition}
\newac{nmf}{NMF}{non-negative matrix factorization}
\newac{umf}{UMF}{unimodality-constrained matrix factorization}
\newac{rmse}{RMSE}{rooted mean squared error}
\newac{olos}{OLOS}{obstructed line-of-sight}
\newac{mmw}{mmW}{millimeter wave}
\newac{ber}{BER}{bit error rate}
\newac{rss}{RSS}{received signal strength}
\newac{lp}{LP}{linear program}
\newac{ufw}{U-FW}{unimodal Frank-Wolfe}
\newac{utf}{UTF}{unimodality-constrained tensor factorization}
\newac{fw}{FW}{Frank-Wolfe}
\newac{iot}{IoT}{Internet-of-Things}
\newac{mae}{MAE}{mean absolute error}
\newac{crb}{CRB}{Cram\'er-Rao bound}
\newac{aoa}{AoA}{angle of arrival}
\newac{wcl}{WCL}{weighted centroid localization}
\newac[plural=GNSS,firstplural=global navigation satellite systems (GNSS)]{gnss}{GNSS}{global navigation satellite system}
\newac{gsm}{GSM}{global system for mobile communications}
\newac{imu}{IMU}{inertial measurement unit}
\newac{rbf}{RBF}{radial basis function}
\newac{msf}{MSF}{multi-sensor fusion}
\newac{lidar}{LiDAR}{light detection and ranging}
\newac{glrt}{GLRT}{generalized likelihood ratio test}
\newac{sdr}{SDR}{software-defined radio}
\newac{ap}{AP}{access point}
\newac{lte}{LTE}{Long-Term Evolution}
\newac{raim}{RAIM}{receiver autonomous integrity monitoring}
\newac{pvt}{PVT}{position, velocity and time}
\newac{npl}{NPL}{Neyman-Pearson lemma}
\newac{ekf}{EKF}{extended Kalman filter}
\newac{tdoa}{TDOA}{time difference of arrival}
\newac{dop}{DOP}{dilution of precision}
\newac[plural=SOPs,firstplural=signals of opportunity (SOPs)]{sop}{SOP}{signal of opportunity}